# Training Beyond Convergence: Grokking nnU-Net for Glioma Segmentation in Sub-Saharan MRI

Mohtady Barakat[1*], Omar Salah[1*], Ahmed Yasser[1], Mostafa Ahmed[1], Zahirul Arief[1], Waleed Khan[1], Dong Zhang[2,3], Aondona Iorumbur[4], Confidence Raymond[3,5,6], Mohannad Barakat[7], Noha Magdy[7]

[1] Faculty of Engineering, Multimedia University, Cyberjaya, Selangor, Malaysia
[2] Department of Electrical and Computer Engineering, University of British Columbia, Vancouver, Canada
[3] Montreal Neurological Institute, McGill University, Montreal, QC, Canada
[4] Department of Physics, Federal University of Technology Minna
[5] Department of Biomedical Engineering, McGill University, Montreal, Canada
[6] Medical Artificial Intelligence Laboratory, Lagos, Nigeria
[7] Department of Computer Science, Friedrich-Alexander-Universität, Erlangen, Germany

barakat.mohtady@gmail.com

**Abstract.** Gliomas are placing an increasingly clinical burden on Sub-Saharan Africa (SSA). In the region, the median survival for patients remains under two years, and access to diagnostic imaging is extremely limited. These constraints highlight an urgent need for automated tools that can extract the maximum possible information from each available scan, tools that are specifically trained on local data, rather than adapted from high-income settings where conditions are vastly different. We utilize the Brain Tumor Segmentation (BraTS) Africa 2025 Challenge dataset, an expert annotated collection of glioma MRIs. Our objectives are: (i) establish a strong baseline with nnUNet on this dataset, and (ii) explore whether the celebrated "grokking" phenomenon an abrupt, late training jump from memorization to superior generalization can be triggered to push performance without extra labels. We evaluate two training regimes. The first is a fast, budget-conscious approach that limits optimization to just a few epochs, reflecting the constrained GPU resources typically available in African institutions. Despite this limitation, nnUNet achieves strong Dice scores: 92.3% for whole tumor (WH), 86.6% for tumor core (TC), and 86.3% for enhancing tumor (ET). The second regime extends training well beyond the point of convergence, aiming to trigger a grokking-driven performance leap. With this approach, we were able to achieve grokking and enhanced our results to higher Dice scores: 92.2% for whole tumor (WH), 90.1% for tumor core (TC), and 90.2% for enhancing tumor (ET).

# 1   Introduction

Brain tumors constitute a significant global health burden, with more than 300,000 new cases and 250,000 deaths recorded worldwide in 2020 alone [1]. Within this group, gliomas represent the second most common primary brain neoplasm, accounting for roughly one-quarter of all cases, and are responsible for the vast majority of malignant presentations [2].

Magnetic Resonance Imaging (MRI) plays a central role in neuro-oncology, offering rich, multiparametric views of brain tumors through sequences like T1-weighted (T1), T1 post- contrast, T2-weighted (T2), and Fluid-Attenuated Inversion Recovery (FLAIR). These scans capture different aspects of tumor structure and behavior, making MRI an essential tool for diagnosis, treatment planning, and monitoring. However, transforming these detailed images into precise, usable measurements still depends heavily on manual effort. Experts often spend over an hour per case carefully outlining tumor regions, and even then, results can vary significantly between individuals [3]. Automated segmentation promises to relieve this burden.

The Brain Tumor Segmentation (BraTS) Challenge has played a pivotal role in advancing machine learning (ML) applications for glioma diagnosis, by providing open-access MRI data, expert annotations, and standardized evaluation metrics. However, it remains uncertain whether ML models trained on BraTS data primarily from high-income settings can generalize to clinical environments in Sub-Saharan Africa [4][5]. Challenges such as fewer annotated cases, lower-resolution imaging, and differing clinical workflows raise important concerns about model transferability [6], In this study, we address these gaps by leveraging the newly released BraTS-Africa dataset [6], designed specifically to reflect the realities of Sub-Saharan clinical imaging.

## 1.1   Related Works

Maruf Adewole [6] implemented an nnU-Net approach for glioma brain tumour segmentation on the BraTS 2021 dataset. Furthermore, Zhao et al. [7] tested whether a BraTS 2021 Challenge (BraTS-GLI 2021) champion-grade nnU-Net method can be adapted to the lower resolution MRIs common in SSA and reflected in the BraTS-Africa Challenge. They compared three regimes: (i) training only on the small BraTS-Africa cohort, (ii) training solely on the 1,251 high-quality BraTS-GLI 2021 cases, and (iii) pre-training on the high-quality data followed by fine-tuning on the African scans. Only the transfer-and-fine-tune strategy generalized well, achieving Dice scores of 0.926 (whole tumor), 0.882 (enhancing tumor) and 0.840 (tumor core). Adding a ×2 super-resolution module did not improve segmentation, which the authors attribute to information lost at acquisition rather than to simple resolution limits [14].

Amod et al. [8] rebuilt a 2022 BraTS runner-up pipeline and evaluated four data setups: Sub-Saharan scans only (BraTS-Africa), high-quality scans only (BraTS-GLI 2021), a merged set of both datasets, and a high-quality model fine-tuned on the BraTS-Africa data. As in Zhao's work, the pre-trained plus fine-tuned variant outperformed models trained solely on the limited low-quality data, which over-fitted, while models trained only on high-quality images under-segmented oedematous tissue in the African scans. The authors conclude that lightweight local adaptation, potentially via federated learning, offers a practical path to robust performance when data sharing is constrained [8].

### 1.2 Model Grokking

Power et al. [10] first described "grokking", a learning pattern in which a network trains for a long time, overfits its dataset, and then, after many more gradient steps, suddenly "gets" the underlying rule and its validation accuracy jumps to nearly 100 % without any change in hyperparameters. In their experiments, the phenomenon suggests that extra training can transform a shaky, memorizing model into a robust one and make it abandon its shortcut solutions. Although the original evidence comes from symbolic data, the idea invites exploration in more complex settings such as brain-tumor segmentation, motivating our decision to push nnU-NetV2 training far beyond conventional stopping points in search of a similar late-phase generalization boost.

### 1.3 nnU-Net and nnU-Netv2

nnU-Net or No New U-net [9] is a model that was created to adapt to the diverse medical datasets and problems available without the need for tailoring a very specific and manual configuration for each problem. By using nnU-Net's automatic configuration creation, non-experts can still use nnU-Net to train on their respective problem and have confidence that it will be among the best accuracies available. nnU- Net does this by the usage of a "fingerprint" extracted from the dataset and a collection of Fixed, Rule-Based and Empirical Parameters chosen at runtime and hardware specification that altogether result in an optimal configuration. For our case, we are using an updated version of nnU-Net, namely nnU-Netv2, for its unified trainer class that we will use to make some slight modifications, cross-platform support of CUDA, MPS, and CPU, flexibility of functionalities using their API calls, and a more user- friendly folder structure.

## 2 Methodology

### 2.1 Study Design and Rationale

For segmentation, we adopted nnU-Netv2 [9] in its three-dimensional (3D) full-resolution configuration. nnU-Net was designed to be flexible on all sorts of medical segmentation tasks which makes it perfect for our case. nnU-Net automatically tailors patch size, depth, and learning schedule to a given dataset, allowing us to concentrate on experimental design rather than hyper-parameter search. We decided to train for 2 different durations, one for 200 epochs and another extended one for 3,500 epochs, for grokking results. Our final hybrid model consists of 2 nnU-Net networks; the main run (200 epochs) an the grokking run (3,500 epochs).

All scans in the BraTS-Africa (SSA) cohort were processed as full 3D volumes. Each input to the model consisted of a 4-channel 3D tensor, where the spatial dimensions (x, y, z) represent the voxel grid and the four channels correspond to the T1, T1 post-contrast, T2, and FLAIR MRI sequences. Look at Fig. 1 for an illustration of such a tensor. This setup allows the model to leverage complementary information across sequences, enabling more accurate and robust tumor segmentation

## 2.2 Dataset

The 2024 BraTS-Africa Challenge dataset was used in this study, consisting of 95 annotated glioma MRI cases, each with four MRI sequences and reference segmentation masks, labelled as the three tumor subregions alongside background (label 0): enhancing tumor (ET, label 1), tumor core (Non-enhancing tumor core) (TC, label 2), and whole tumor (Surrounding non-enhancing FLAIR hyperintensity) (WT, label 3) [14-15]. Sixty cases were designated for training and internal validation. Of these, five were randomly withheld before processing for internal validation, and the remaining 55 were used for all model training. For the pre-submission evaluation, predictions were submitted for 35 validation cases provided without ground truth labels. All reported performance metrics are based on BraTS's evaluation of these 35 cases and the final testing dataset from the BraTS-Africa Challenge 2025.

## 2.3 Preprocessing Pipeline

All raw NIfTI volumes were used without manual modification. Key dataset properties such as voxel spacing, foreground intensity ranges, and median shape size were then analyzed to define a common resampling target and optimize patch and batch sizes for training. All images and corresponding segmentation masks were resampled to the same voxel spacing to ensure uniform physical resolution across the dataset.

Each MRI sequence was normalized independently using z-score normalization: brain voxels were centered by subtracting the mean and scaled by the standard deviation, while non-brain voxels retained a zero value to clearly distinguish background. This approach enables consistent intensity scaling across patients and sequences without the need for manual adjustments.

Training included a comprehensive set of spatial and intensity augmentations applied on-the-fly. These included random rotations, scaling, flipping, elastic deformations, brightness and contrast changes, gamma adjustments, Gaussian noise, and simulated low-resolution artifacts. Occasional small components were also removed from the label maps to improve robustness. The input data consisted of 3D voxel patches with all four MRI modalities stacked as channels and processed jointly for segmentation.

## 2.4 Training Configuration

We adopted nnU-Net v2 in its 3D full-resolution preset and left every internal hyperparameter exactly as the framework generated it. Once the training data were analyzed, nnU-Net's heuristic planner produced a six-stage U-Net shown in Figure 1 with an initial feature width of 32, deep-supervision, and an input-patch size of 128 × 160 × 112 voxels that comfortably fit into the 22.5 GB L4 memory. The optimizer remained the default stochastic gradient descent with Nesterov momentum ($\mu = 0.99$), a weight- decay of $3 \times 10^{-5}$, and a polynomial learning rate decay starting at 0.01 and annealing to zero by the final epoch as can be seen in Figure 2. The number of iterations per epoch is 300, 50 of them being for validation. The number of input channels is based on the number of modalities, which is 4.

Loss was the standard Dice + cross-entropy combination applied to each deep-supervision output, and mixed-precision training was enabled via NVIDIA's AMP. No manual overrides, layer additions, or schedule tweaks were introduced at any point; the configuration is therefore fully reproducible by invoking nnU-Net v2 with the same dataset and random seed. This was trained for 200 epochs. We also trained another version with the same configuration, but we trained for 3500 epochs for grokking.

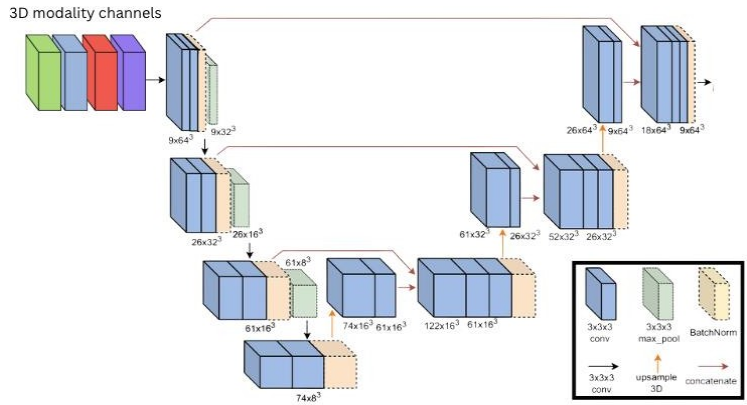

**Figure 1.** Illustration of 3D voxel batches with modalities overlaid and input into the 3D Unet at full resolution

### 2.5 Training Baseline run (200 epochs)

The first experiment trained the network for 200 epochs on the 55-scan set. Batches were shuffled per epoch. Training was executed on a single NVIDIA L4 (22.5 GB) GPU, taking ≈ 100s per epoch (wall-clock ≈ 6 h total).

### 2.6 Training Extended Grokking run (3500 epoch)

Motivated by recent observations that some networks "grok" (i.e., memories training data early yet require far longer to internalize the underlying rule set), we trained a model for 3,500 epochs. The learning rate scheduler was re-initialized to its starting value, and it was limited to a minimum value of $1 \times 10^{-4}$ as shown in Figure 2A. The average epoch time remained 100s, leading to an uninterrupted runtime of roughly five days. The loss progress for the start of the training compared with after grokking is shown in Fig. 3.

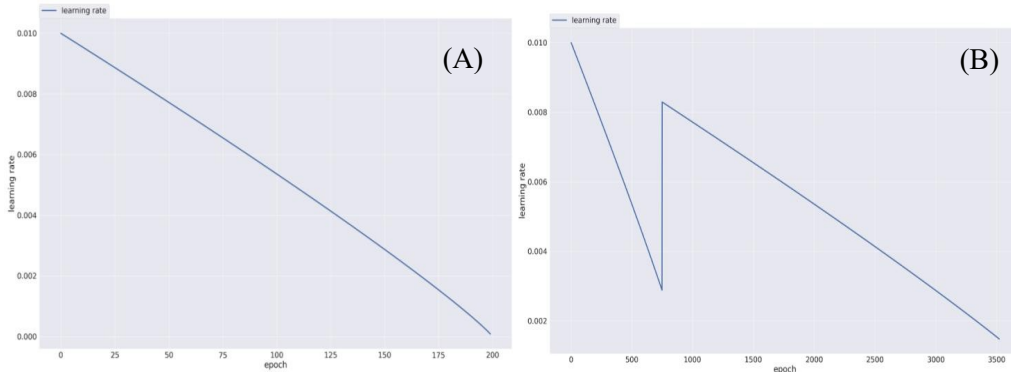

**Figure 2.** ( A ) Learning rate graph of a 200-epoch baseline run. (B) Learning rate graph of 3,500 epoch Extended Grokking run.

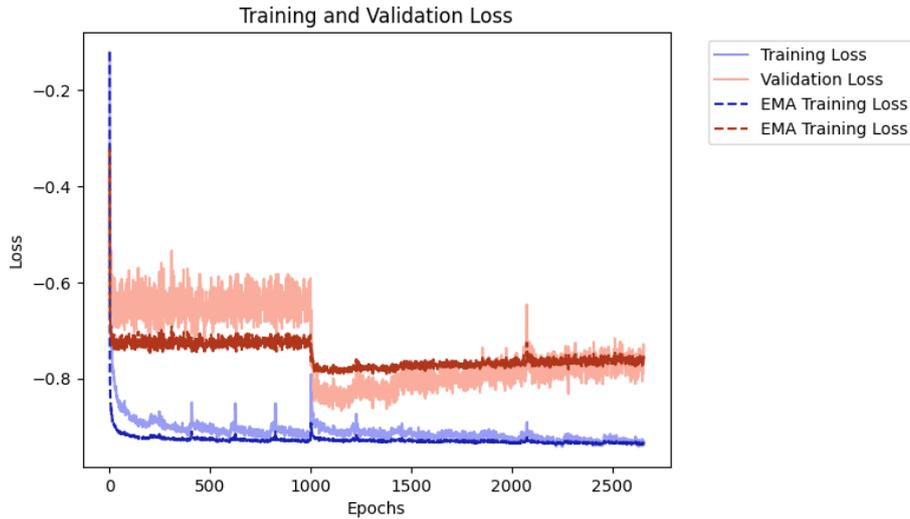

**Figure 3.** Loss change over the training epochs. Training in 100 epochs, overfitting in 1000 epochs, and grokking after 1000 epoch

## 3   Results and Discussion

### 3.1   Main Run (200 Epoch)

The original nnU-Net recipe (full-resolution 3D) was quantitatively evaluated on the validation set of 35 BraTS MRI cases. Performance was assessed as the mean scores of the following two metrics across all 35 cases:

- Dice Coefficient: how much the predicted mask volume overlaps the ground truth.
- Normalised Surface Dice (NSD): how closely the surfaces match within a chosen tolerance.

Both metrics are shown per case (one score per subject) and Lesion-Wise (LW: each connected lesion is treated as a separate object; the metric is computed per lesion then averaged). Table 1 below presents relevant scores for the three relevant tumor sub-regions/classes. Note that NSD is computed for 1mm and 0.5mm and shown in Fig. 4.

*Table 1.* Mean scores for the three relevant tumor classes with standard deviation (for LW) for the main run (200 epochs)

| Metric | Enhancing Tumor | Tumor Core | Whole Tumor |
| --- | --- | --- | --- |
| Dice coefficient | 0.866 | 0.863 | 0.923 |
| LW Dice | 0.821±0.257 | 0.816±0.272 | 0.880±0.208 |
| NSD 0.5 mm | 0.586 | 0.511 | 0.552 |
| NSD 1.0 mm | 0.851 | 0.777 | 0.836 |
| LW NSD 0.5 mm | 0.560±0.219 | 0.498±0.221 | 0.522±0.178 |
| LW NSD 1.0 mm | 0.805±0.266 | 0.736±0.274 | 0.789±0.207 |

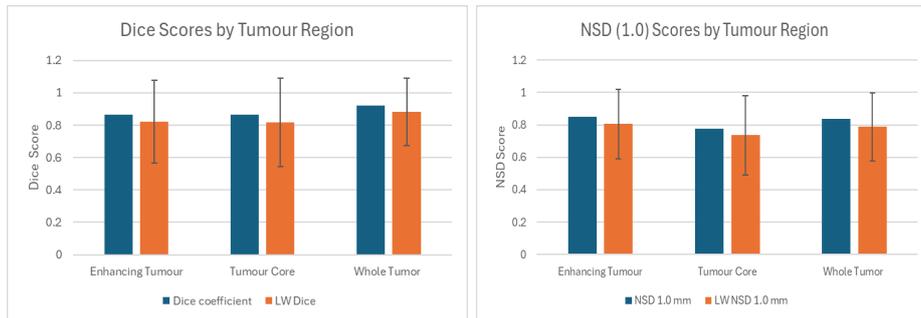

**Figure 4.** Mean Dice scores and normalized surface Dice (NSD) at 1.00mm (including lesion-wise) by tumor region for the main run (200 epochs).

As illustrated in Figure 5, in every slice the model captures the three tumor sub-regions much like the ground-truth masks. But, in some cases it can overlook small Lesions. These scenes match the overall trend: the network is good at finding the bulk of each tumor but still misses some edges and tiny parts of the tumors

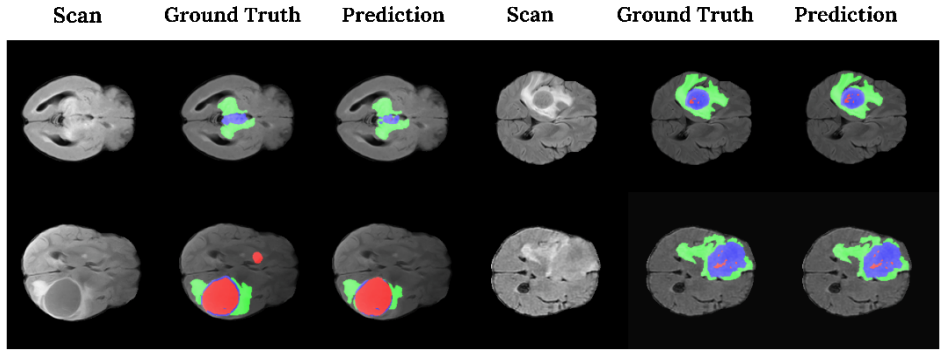

**Figure 5.** An illustration of four different cases, each showing the acquired scans, the ground-truth mask, and our nnU-Net predictions for all three tumor sub-regions (red = tumor core, blue = enhancing tumor and the green = whole tumor.

### 3.2 Grokking Run (3500 epoch)

The extended nnU-Net run, now trained for 3,500 epochs to explore grokking-style dynamics, was quantitatively evaluated on the same 35 cases. Each metric is reported per subject and in a lesion-wise (LW) view, where every connected tumor focus is treated as a separate object before averaging; the LW view therefore penalizes missed satellites and split errors more strongly than the per-case aggregate. Table 2 presents relevant scores for the 3 relevant classes. Additionally, Figure 6 below represents the grouped bar charts with the same figures.

*Table 2.* Mean scores for the 3 relevant classes with standard deviation (for LW) for the grokking run (3,500 epochs)

| Metric | Enhancing Tumor | Tumor Core | Whole Tumor |
| --- | --- | --- | --- |
| Dice coefficient | 0.902 | 0.901 | 0.922 |
| LW Dice | 0.860±0.201 | 0.846±0.228 | 0.874±0.216 |
| NSD 0.5 mm | 0.622 | 0.551 | 0.541 |
| NSD 1.0 mm | 0.893 | 0.816 | 0.828 |
| LW NSD 0.5 mm | 0.582±0.208 | 0.520±0.225 | 0.513±0.179 |
| LW NSD 1 mm | 0.837±0.216 | 0.764±0.244 | 0.780±0.210 |

Comparing both networks shows that Grokking clearly benefits the enhancing tumor and tumor-core regions: Dice improves from 0.866 to 0.902 for ET and from 0.863 to 0.901 for TC, indicating sharper delineation of the clinically most critical compartments. For whole tumor, however, the 200-epoch model matches the 3 500-epoch version (0.923 versus 0.922).

These complementary strengths imply our hybrid model strategy: deploying the 3500-epoch network to segment ET and TC, while retaining the 200-epoch model for WT to yield the best overall segmentation.

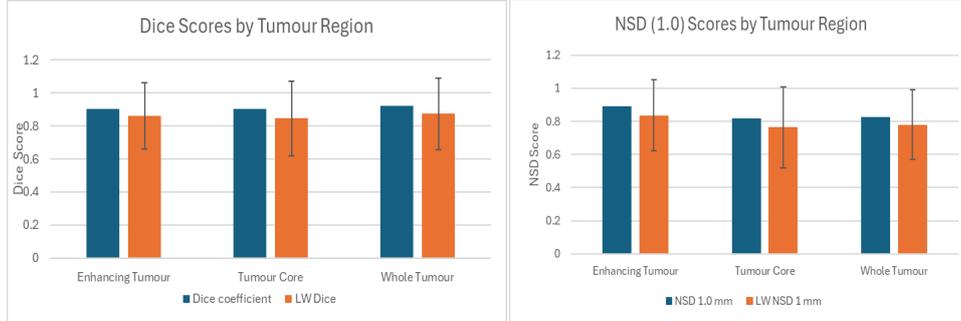

**Figure 6.** Grouped bar charts for Dice scores by tumor region & NSD (1.0mm) by tumor region for the grokking run (3,500 epochs)

### 3.3 Hybrid Model (200 & 3500 epoch networks)

We tested and validated the hybrid model network on the BraTS-Africa 2025 validation cases (35 cases) and the BraTS-Africa 2025 test cases. Table 3 presents relevant scores for the 3 relevant classes. Additionally, Figure 7 below represents the grouped bar charts with the same figures.

*Table 3.* Mean scores for the 3 relevant classes with standard deviation (for LW) for the hybrid network (3,500 & 200 epoch models)

| Metric | Enhancing Tumor | Tumor Core | Whole Tumor |
|---|---|---|---|
| Dice coefficient | 0.902 | 0.903 | 0.922 |
| LW Dice | 0.859±0.182 | 0.831±0.226 | 0.864±0.209 |
| NSD 0.5 mm | 0.622 | 0.551 | 0.541 |
| NSD 1.0 mm | 0.891 | 0.817 | 0.829 |
| LW NSD 0.5 mm | 0.595±0.187 | 0.516±0.212 | 0.508±0.165 |
| LW NSD 1 mm | 0.852±0.180 | 0.761±0.217 | 0.778±0.182 |

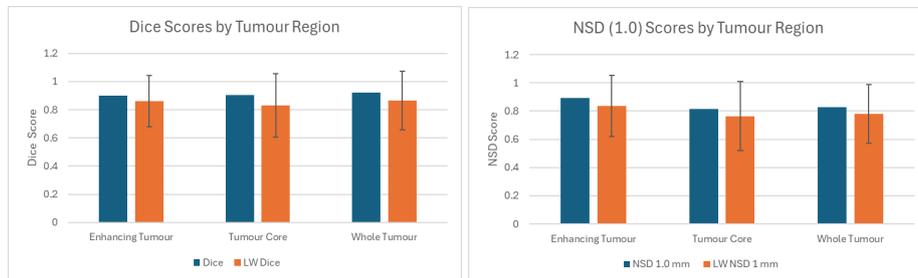

**Figure 7.** Bar charts for Dice scores by tumor region & NSD (1.0mm) by tumor region for the hybrid model (3500-epoch network to segment ET and TC and the 200-epoch model for WT)

### 3.4 Test cases

Running our model on the test cases dataset (Test cases of BraTS-Africa 2025) our model performed better as shown in table 4.

*Table 4.* Mean scores for the 3 relevant classes with standard deviation (for LW) for the hybrid network (3,500 & 200 epoch models) on test cases

| Metric | Enhancing Tumor | Tumor Core | Whole Tumor |
|---|---|---|---|
| LW Dice | 0.854±0.173 | 0.858±0.181 | 0.913±0.147 |
| LW NSD 1 mm | 0.866±0.173 | 0.810±0.196 | 0.848±0.133 |

## 4 Conclusion

This study evaluated nnU-Netv2 on the BraTS-Africa 2025 dataset to establish a baseline for glioma segmentation on African MRI data and to explore whether prolonged training could trigger the "grokking" phenomenon. A standard 200-epoch run already achieved strong Dice and NSD scores, while extending training to 3,500 epochs further improved enhancing tumor and tumor core. Whole tumor performance remained largely unchanged, suggesting a trade-off where longer training sharpens critical subregions. These results provide a reproducible baseline and highlight resource-aware strategies for improving automated segmentation in Sub-Saharan clinical settings.

### Acknowledgements

This work was part of the Sprint AI Training for African Medical Imaging Knowledge Translation (SPARK) Academy 2025 summer school on deep learning in medical imaging. The authors would like to thank the instructors of the summer for providing insightful background knowledge on brain tumours that informed the research presented here, most notably, Noha Magdy, Maruf Adewole, Ayomide B. Oladele, Amal Saleh, Nourou Dine Bankole, Jeremiah Fadugba, Teresa Zhu, Craig Jones, Charles Delahunt, Celia Cintas, Lukman E. Ismaila, Ugumba Kikwima, Mehdi Astaraki, Pe- ter Hastreiter, Evan Calabrese, Esin Uzturk Isik, , Rancy Chepchirchir, James Gee, MacLean Nasrallah, Jean Baptiste Poline, Bijay Adhikari, Kenneth Agu, Mohannad Barakat & Yahoo Liu. The authors acknowledge the computational infrastructure support from the Digital Research Alliance of Canada (The Alliance) and the University of Washington Azure GenAI for Science Hub through the eScience Institute and Microsoft (PI: Mehmet Kurt) secured for the SPARK Academy. Finally, we thank the Lacuna Fund for Health and Equity (PI: Udunna Anazodo, 0508-S-001), the RSNA R&E Foundation (PI: Farouk Dako), McGill Heatly Brain and Healthy Lives (HBHL; Udunna Anazodo), as well as the National Science and Engineering Research Council of Canada (NSERC) Discovery Launch Supplement (PI: Udunna Anazodo, DGECR.